\documentclass[12pt]{article}

\begin{document}

\begin{titlepage}

\begin{flushright}
IUHET-491\\
\end{flushright}
\vskip 2.5cm

\begin{center}
{\Large \bf Eliminating the $CPT$-Odd $f$ Coefficient from the Lorentz-Violating
Standard Model Extension}
\end{center}

\vspace{1ex}

\begin{center}
{\large B. Altschul\footnote{{\tt baltschu@indiana.edu}}}

\vspace{5mm}
{\sl Department of Physics} \\
{\sl Indiana University} \\
{\sl Bloomington, IN 47405 USA} \\

\end{center}

\vspace{2.5ex}

\medskip

\centerline {\bf Abstract}

\bigskip

The fermionic $f$ coefficient in the Lorentz-violating standard model extension
presents a puzzle. Thus far, no observable quantity that depends upon $f$
has ever been found.
We show that this is because $f$ is actually unnecessary. It has absolutely no
effects at leading order and can be completely absorbed into other
coefficients of the theory by a redefinition of the field.

\bigskip

\end{titlepage}

\newpage

In recent years, there has been a growing interest in the possibility that there
could exist small Lorentz- and $CPT$-violating corrections to the standard
model. A number of candidate theories of quantum gravity predict possible
violations of these fundamental symmetries, and if any such violations were
found, they would be important clues regarding the nature of  Planck scale
physics. An effective field theory, the standard model extension (SME) has been
developed to describe all possible violations of Lorentz symmetry in
quantum field theory~\cite{ref-kost1,ref-kost2} and gravity~\cite{ref-kost5}.
The full SME is quite complicated, and so we typically restrict our attention to
a field theory with only a finite number of Lorentz-violating parameters.
The minimal SME contains only operators that are superficially renormalizable,
and both the one-loop renormalization~\cite{ref-kost4} and the
stability~\cite{ref-kost3} of this theory have been studied.

To date, experimental tests of Lorentz violation
have included studies of matter-antimatter asymmetries for
trapped charged particles~\cite{ref-bluhm1,ref-bluhm2,ref-gabirelse,
ref-dehmelt1} and bound state systems~\cite{ref-bluhm3,ref-phillips},
determinations of muon properties~\cite{ref-kost8,ref-hughes}, analyses of
the behavior of spin-polarized matter~\cite{ref-kost9,ref-heckel},
frequency standard comparisons~\cite{ref-berglund,ref-kost6,ref-bear},
measurements of neutral meson oscillations~\cite{ref-kost7,ref-hsiung,ref-abe},
polarization measurements on the light from distant galaxies~\cite{ref-carroll1,
ref-carroll2,ref-kost11}, and others. The results of these tests can be used to
place bounds on many of the minimal SME's Lorentz-violating coefficients.
However, there are still many sectors of the theory for which there are no
useful bounds at all. Since the minimal SME is used to parameterize the possible
forms of Lorentz violation that might be seen in experiments, it is important to
understand the structure of the model itself. In particular, we should know how
many independent forms of Lorentz violation the theory can describe.

In the course of analyzing these many tests of Lorentz symmetry, one puzzling fact
has been observed. One set of Lorentz-violating coefficients in the Lagragian---the
$f$ terms in the fermion sector---always
seem to cancel out when we calculate observable quantities.
(As an immediate consequence, there are no known experimental bounds on any $f$.)
In this paper, we shall look more closely at these coefficients. We shall
show that the lack of any leading-order experimental dependences
on $f$ is actually a natural consequence of its
structure. In fact, $f$ can be completely eliminated from any theory by
redefining the fields. The $f$ is reabsorbed into a different Lorentz-violating
parameter, and the lowest-order $f$-dependent effects are of second order in the
Lorentz violation. This
means that $f$ is entirely unnecessary to our description of the theory, and we
may dispense with it entirely (unless using
it happens to be convenient in a particular situation). These results resolve
a significant puzzle, and they result in a valuable reduction in the complexity
of the minimal SME.

We must begin by introducing the theory.
For a model with a single species of fermion, the most general superficially
renormalizable SME Lagrange density is
\begin{equation}
\label{eq-L}
{\cal L}=\bar{\psi}(i\Gamma^{\mu}\partial_{\mu}-M)\psi,
\end{equation}
where
\begin{equation}
\Gamma^{\mu}=\gamma^{\mu}+c^{\nu\mu}\gamma_{\nu}-d^{\nu\mu}\gamma_{\nu}
\gamma_{5}+e^{\mu}+if^{\mu}\gamma_{5}+\frac{1}{2}g^{\lambda\nu\mu}
\sigma_{\lambda\nu}.
\end{equation}
and
\begin{equation}
M=m+\!\not\!a-\!\not\!b\gamma_{5}+\frac{1}{2}H^{\mu\nu}\sigma_{\mu\nu}+im_{5}
\gamma_{5},
\end{equation}
Some of the coefficients
in ${\cal L}$ are more important than others. For example, $m_{5}$ is
not Lorentz violating, and it may be absorbed into the
other coefficients by means of a particular field
redefinition,
\begin{equation}
\label{eq-m5}
\psi'=e^{-\frac{i}{2}\gamma_{5}\tan^{-1}(m_{5}/m)}\psi,
\end{equation}
that was already known before the introduction
of the SME. This and other field redefinitions are discussed in detail
in~\cite{ref-colladay2}, although only up to leading order. In this paper, we
shall be looking at effects of field redefinitions beyond leading order as well.

An $a$ term can also be completely eliminated from the single-fermion theory,
since it is essentially nothing more than a constant classical vector potential
term. Removing $a$ simply redefines the origin in momentum space---$p\rightarrow
p-a$. However, if there are multiple species and flavor-changing interactions,
differences in their $a$ terms can be observable, and gravitational effects could
also make $a$ an observable quantity.

Slightly different is the antisymmetric part of $c$, $c^{[\nu\mu]}=c^{\nu\mu}-
c^{\mu\nu}$.
At leading order, the
$c^{[\nu\mu]}$ terms are equivalent to a redefinition of the Dirac matrices;
such a rotation in spinor space
can have no physical effects. So this part of $c$ can be eliminated with
another field redefinition, but only if the ${\cal O}(c^{2})$ terms are neglected.
That the antisymmetric terms do contribute at higher order is evident from the
fermions' energy-momentum relation, which is given below as
equation~(\ref{eq-Ec}).

There are also other reasons to believe that some coefficients may be more
interesting than others. The $e$, $f$, and $g$ kinetic couplings
appear superficially inconsistent
with the coupling of the fermion field to standard model gauge fields, because
the mix left- and right-chiral fields. Such
terms could only arise at the electroweak breaking scale, as vacuum expectation
values of nonrenormalizable operators, and so they might then be expected to be
less important than the $c$ and $d$ terms. [However, as we shall see, there is
a significant weakness in this argument. We are using the conventional
Lorentz-invariant definition of the chirality operator, which might not be
appropriate when the $SU(2)_{L}$ gauge group is coupled to Lorentz-violating
matter.]

The coefficient $f$ is similar to both $a$ and the antisymmetric part of $c$
Like $a$, $f$ is completely unnecessary for describing
the physics. However, unlike $a$, $f$ has definite physical effects, although
only beyond leading order.  What makes $f$ superfluous in the formalism is not
that this term is unphysical, but that the effects it generates are exactly the
same as those generated by another Lorentz-violating term. The more general $c$
subsumes all the physics of an $f$ coefficient, and the $f$ can be eliminated by
absorbing it into $c$. At second order in $f$, the effects of $f$ are
indistinguishable from those of a $c$ term
\begin{equation}
c^{\nu\mu}=-\frac{1}{2}f^{\nu}f^{\mu}.
\end{equation}
This situation is also similar to what occurs with $m_{5}$, as each of these terms
can be entirely absorbed into other coefficients in the theory.

So far, there are no experimental bounds on $f$. In fact, of all the
Lorentz-violating coefficients $M$ and $\Gamma^{\mu}$, $f$ is the only one for
which there are currently no suggestions even for how it might be bounded.
Typically, when searching for experimental tests of Lorentz violation, we
restrict our attention to effects that appear at first order in the coefficients.
Because Lorentz violation is small, any higher-order effects should be miniscule
and would only be important if they caused a qualitative change in the structure
of a theory. (For example, at second order in $b$, radiative corrections
to QED could violate gauge invariance and possibly lead to a photon
mass~\cite{ref-altschul1,ref-altschul2}.)
The $f$ coefficient has no physical effects at leading order, and that is
precisely why its value is not constrained.

The fact that $f$ has no leading-order effects on a theory is also related to the
discrete symmetries associated with this operator. The timelike coefficient
$f^{0}$ is separately odd under $C$, $P$, and $T$. These are
the same symmetries as are possessed by the spacelike parts of $a$ and $e$.
However, there are other discrete symmetries that distinguish these operators.
The parity operator is defined as inverting all the spatial coordinates,
$\vec{x}\rightarrow-\vec{x}$. However, $P$ may be broken down into the product of
three separate reflections, $P=R_{1}R_{2}R_{3}$, where $R_{j}$ takes $x_{j}
\rightarrow-x_{j}$ and leaves the other two coordinates unchanged. While
$a_{j}$ and $e_{j}$ (for fixed $j$) are odd under $R_{j}$, they are even under
the other two
reflections. However, $f_{0}$ is odd under all the $R_{j}$. No other minimal
SME coefficient has this property. These curious symmetry properties mean that
there is no other object in the theory that can combine with $f$ to give, for
example, something with the form of an ${\cal O}(f)$ energy shift. For
similar reasons, $f$ does not mix with any other coefficients
under the action of the renormalization group~\cite{ref-kost4}. (In fact, there
are not even any self-renormalization terms in the one-loop
$\beta$-function for $f$; $\beta_{f}$ vanishes identically at leading order.)

Now to see the plausibility of our main claim, that any $f$ term can be absorbed
into $c$, let us look at the energy-momentum
relation separately in the presence of purely spacelike $c$ and $f$ coefficients.
The energies then are
\begin{equation}
\label{eq-Ec}
E=\sqrt{m^{2}+\left(p_{k}-c_{kj}p_{j}\right)\left(p_{k}-c_{kl}p_{l}
\right)}
\end{equation}
and
\begin{equation}
E=\sqrt{m^{2}+p_{j}p_{j}+\left(f_{k}p_{k}\right)^{2}}.
\end{equation}
These are actually very similar. Note that if there is only a $c_{33}$ or an
$f_{3}$, then each of these dispersion relations takes the form
\begin{equation}
E=\sqrt{m^{2}+p_{1}^{2}+p_{2}^{2}+\xi p_{3}^{2}},
\end{equation}
where
$\xi$ is either $(1-c_{33})^{2}$ or $1+f_{3}^{2}$. This is sufficient to show
that the noninteracting theories with purely spacelike $c$ and $f$ are equivalent.
However, we obviously want to show more---that this equivalence can continue even
in more complicated situations.

For definiteness, we shall continue to work with a theory containing $f_{3}$ only,
demonstrating how this may be transformed into a $c_{33}$. This can then be
generalized to cover other cases without too much difficulty, although there are
some additional subtleties that arise when a timelike $f$ is considered. The
Lagrange density with $f_{3}$ only reduces to
\begin{equation}
{\cal L}= \bar{\psi}\left(i\!\not\!\partial-\gamma_{5}f_{3}\partial_{3}-m
\right)\psi.
\end{equation}
Everything is conventional, except for the matrix multiplying $\partial_{3}$.
The usual $\gamma_{3}\partial_{3}$ has been replaced by $\left(\gamma_{3}-if_{3}
\gamma_{5}\right)\partial_{3}$.

The crucial observation is that $\gamma_{3}-if_{3}\gamma_{5}$ anticommutes with
$\gamma^{\mu}$ for $\mu\neq3$, just as does $\gamma_{3}$ itself. The matrices
$\gamma_{3}$ and $i\gamma_{5}$ are actually completely interchangeable in the
ordinary Dirac theory; they satisfy exactly the same anticommutation relations
with the other Dirac matrices and possess the same normalization. So any
$\gamma_{3}\cos\theta-i\gamma_{5}\sin\theta=\gamma_{3}e^{i\gamma_{3}\gamma_{5}
\theta}$ can actually be substituted for $\gamma_{3}$ in the Lorentz-invariant
Lagrangian without affecting the physics.

However, $\gamma_{3}-if_{3}\gamma_{5}$ does not quite have this form. Instead,
\begin{equation}
\gamma_{3}-if_{3}\gamma_{5}=\sqrt{1+f_{3}^{2}}\gamma_{3}e^{i\gamma_{3}\gamma_{5}
\tan^{-1}f_{3}}.
\end{equation}
The rescaling factor $\sqrt{1+f_{3}^{2}}$ gives rise to the nontrivial $c_{33}$
at ${\cal O}(f^{2})$. Defining new $\gamma$-matrices by
\begin{equation}
\gamma'_{\mu}=\left\{
\begin{array}{ll}
\gamma_{\mu}, & \mu\neq3 \\
\gamma_{3}e^{i\gamma_{3}\gamma_{5}\tan^{-1}f_{3}}, & \mu=3
\end{array}
\right.
\end{equation}
transforms the Lagrange density into
\begin{equation}
{\cal L}=\bar{\psi}\left[i\gamma'_{\mu}
\partial^{\mu}-i\left(\sqrt{1+f_{3}^{2}}-1\right)\gamma'_{3}\partial_{3}-m\right]
\psi
\end{equation}
---that is,
a Lagrange density for a theory with a $c_{33}=1-\sqrt{1+f_{3}^{2}}$ only.

The generalization to an arbitrary purely spacelike $f$ is elementary:
\begin{equation}
\gamma'_{j}=\gamma_{j}e^{if_{k}\gamma_{k}\gamma_{5}G
\left(f_{l}f_{l}\right)}=e^{-\frac{i}{2}f_{k}\gamma_{k}\gamma_{5}G
\left(f_{l}f_{l}\right)}\gamma_{j}e^{\frac{i}{2}f_{k}\gamma_{k}\gamma_{5}G
\left(f_{l}f_{l}\right)},
\end{equation}
where $G(x)=\frac{1}{\sqrt{x}}\tan^{-1}\sqrt{x}$. Note that $G(x)$ is analytic
around $x=0$, so the arguments of the exponents depend analytically on the
components of $f$. It is a trivial matter to recast this as a redefinition of
the field, rather than the Dirac matrices. The exponentials commute with
$\gamma_{0}$, so the field redefinition is simply
\begin{eqnarray}
\label{eq-redefspace}
\psi' & = & e^{-\frac{i}{2}f_{k}\gamma_{k}\gamma_{5}G\left(f_{l}f_{l}\right)}\psi
\\
\bar{\psi}' & = & \bar{\psi}
e^{\frac{i}{2}f_{k}\gamma_{k}\gamma_{5}G\left(f_{l}f_{l}\right)},
\end{eqnarray}
and the Lagrangian for $\psi'$ contains only a $c$ term.

What is left is to deal with timelike $f_{0}$ terms. The correct generalization of
(\ref{eq-redefspace}) is obvious:
\begin{equation}
\label{eq-redefgeneral}
\psi'=e^{\frac{i}{2}f^{\mu}\gamma_{\mu}\gamma_{5}G\left(-f^{2}\right)}\psi;
\end{equation}
but there are some slight
complications associated with the timelike case. For a purely timelike $f$, 
with $f_{0}$ only, the field redefinitions become
\begin{eqnarray}
\label{eq-redeftime}
\psi' & = & e^{\frac{i}{2}\gamma_{0}\gamma_{5}\tanh^{-1}f_{0}}\psi
\\
\bar{\psi}' & = & \bar{\psi}e^{-\frac{i}{2}\gamma_{0}\gamma_{5}\tanh^{-1}f_{0}}.
\end{eqnarray}
This converts $f_{0}$ into a $c_{00}=\sqrt{1-f_{0}^{2}}-1\approx-\frac{1}{2}
f_{0}^{2}$, and here we see the subtlety. For spacelike $f$, the transformation
could be effected for an arbitrary negative $f^{2}$; however, when the
Lorentz-violating coefficient is timelike, the field redefinition is only possible
if $f^{2}<1$. Larger values give rise to an imaginary $c$ and so a Lagrangian that
is not Hermitian. This is not surprising, for if
$f^{2}>1$, then the square of the matrix multiplying
$\partial_{0}$ in the Lagrangian becomes negative, and so the time evolution can
no longer be unitary. In that case, the entire theory becomes inconsistent.

Since the field redefinition (\ref{eq-redefgeneral}) works to eliminate $f$ in
both the purely spacelike and purely timelike cases, it can be shown to hold
for an arbitrary $f$, simply by performing the relevant calculations in a boosted
frame. The $c$ that is generated by the transformation is
\begin{equation}
\label{eq-c}
c^{\nu\mu}=\frac{f^{\nu}f^{\mu}}{f^{2}}\left(\sqrt{1-f^{2}}-1\right).
\end{equation}
This may be further generalized to the case in which the initial Lagrangian has
both $c$ and $f$ terms. In that case, the $f$ may still be absorbed into a
modification of $c$, and again there are no ${\cal O}(f)$ terms. However, the
resulting expression is rather cumbersome, and it is uninteresting practically.
What is important is that the leading contribution that $f^{\mu}$ makes to
$c^{\nu\mu}$ is
unchanged and remains equal to $-\frac{1}{2}f^{\nu}f^{\mu}$.
Infinitesimally, the field redefinition (\ref{eq-redefgeneral}) we have found is
identical with that presented in~\cite{ref-colladay2}, where it was pointed out
that this would eliminate $f$ from the free Lagrangian at leading order.
Also as discussed in~\cite{ref-colladay2}, this kind of transformation will
generally reshuffle any other Lorentz-violating coefficients that are present in
the theory. For example, if the theory prior to the elimination of $f$ contains
a $b$ term, then the field redefinition will generate a $H^{\mu\nu}$ proportional
to $(f^{\mu}b^{\nu}-b^{\mu}f^{\nu})$. Likewise, if the theory initially contains
an $a$, the field redefinition will generate a $m_{5}$ proportional to $f^{\mu}
a_{\mu}$; fortunately however, if there exists a concordant frame, in which all
the Lorentz-violating coefficients are small, it is indeed possible to eliminate
both $f$ and $m_{5}$ from the theory, using slightly more involved field
redefinitions.

The expression (\ref{eq-c}) is indeterminate for lightlike $f$, but the
limiting value as $f^{2}\rightarrow0$, $c^{\nu\mu}=-\frac{1}{2}f^{\nu}f^{\mu}$,
is correct
at $f^{2}=0$. This can be verified, for example, using light cone coordinates.
Otherwise, (\ref{eq-c}) holds formally for all other $f^{2}<1$ [although, as with
$G(-f^{2})$, the power series expansion about $f^{2}=0$ fails for $f^{2}<-1$].
However, the larger-$f$ behavior of the theory is fairly uninteresting, for two
reasons. First, for all observed particles, Lorentz violation is small. Second, if
a theory did contain a large $f$ or large $c$, there would be causality
violations at a low energy scale, invalidating the description in terms of
effective field theory anyway~\cite{ref-kost3}.

We expect the coefficients describing any
physical Lorentz violation to be of  characteristic size
${\cal O}(m/M_{P})$, where $m$ is a typical mass scale (i.e., in the $\sim$1--100
GeV range), and $M_{P}$ is some very large scale, possibly the Planck scale.
Typically, the description of the physics in terms of effective field theory
breaks down at energies comparable to $M_{P}$. Additional higher-dimension
operators must be introduced at that scale if properties such as causality are to
be preserved. However, the $c$ term is an exception to this. Because $c$ and the
Lorentz-invariant kinetic term possess the same basic structure, there is mixing
between them, and the $c$-modified theory fails at the lower scale
$\sqrt{mM_{P}}$.

To see this, observe that the velocity in the presence of a purely spacelike $c$
(chosen for simplicity) is
\begin{equation}
\label{eq-v}
v_{k}=\frac{1}{E}\left(p_{k}-c_{kj}p_{j}-c_{jk}p_{j}+c_{jk}c_{jl}p_{l}
\right).
\end{equation}
This can become superluminal when $|\vec{p}|/E
\approx 1-|c|$, where $|c|$ is a characteristic size for the Lorentz-violating
coefficients.
For ultrarelativistic particles, for which $1-|\vec{v}\,|\ll 1$, the 
Lorentz factor is roughly $\gamma\approx1/\sqrt{2(1-|\vec{v}\,|)}$. This sets the
scale of $\gamma$ at which new physics must enter:
$\gamma_{\max}\sim1/\sqrt{|c|}$. This corresponds to an
energy scale $E_{\max}\sim\sqrt{mM_{P}}$.

So it seems there may be a conflict between the version of the theory containing
$f$, which breaks down at the higher scale $M_{P}$, and the version with $c$,
which could fail at a lower scale. However, this problem is alleviated by the
fact that the $c$ term related to $f$ is actually of ${\cal O}(f^{2})$, and so
its natural size is ${\cal O}(m^{2}/M^{2}_{P})$. When $f$ is converted into $c$,
the energy at which things
break down is just the geometric mean between $m$ and $M_{P}^{2}/m$, and this
is exactly $M_{P}$. So the scale of new physics is defined consistently in either
framework.

The fact that $f$ can be absorbed into $c$ is also related to the leading order
triviality of $c^{[\nu\mu]}$. There are five independent
mutually anticommuting $4\times4$ matrices, which may be arranged in any way we
like as the $\gamma^{\mu}$ and $\gamma_{5}$ (with appropriate factors of $i$).
The elimination of $f$ fixes the
definition of $\gamma_{5}$ and removes four of the ten degrees of freedom
associated with changes to the representations of the Dirac matrices. However,
there are still six unphysical degrees of freedom contained in $c$. The
quantity $g^{\nu\mu}+c^{\nu\mu}$ defines a bilinear form that connects
$p_{\mu}$ and $\gamma_{\nu}$ in the action. This bilinear form contains sixteen
free parameters. However, the physics in a theory with a $c$-type modification
ultimately depends only on
energy-momentum relation, which can be expressed as a bilinear form that connects
$p$ with itself. So only the symmetric part of this second bilinear form is
physical, and this amounts to only ten physical parameters. The six parameters
that are unphysical are exactly those that correspond to the $SO(3,1)$
transformations that change the representation of the $\gamma^{\mu}$. At leading
order, these transformations are represented precisely by $c^{[\nu\mu]}$;
however, at higher orders, the
algebraic characterization of which parts of $c$ are trivial
becomes more complicated.

However, there is a
fairly simple geometrical characterization of which parts of $c$ are actually
physical. With $c$ as the only form of Lorentz violation, the fermionic
energy-momentum relation takes the general form
\begin{equation}
\label{eq-EMwithC}
C^{\nu\mu}C_{\nu}\,^{\rho}p_{\mu}p_{\rho}-m^{2}=0,
\end{equation}
in terms of $C^{\nu\mu}=g^{\nu\mu}+c^{\nu\mu}$. We shall work in a fixed frame
and consider $C^{\nu\mu}=(C^{\nu})^{\mu}$ as a ``vector of vectors.'' The
inner index ($\nu$) is coupled to the specific Dirac representation, while the
outer index ($\mu$) may be seen simply as a parameter. It is then clear from
(\ref{eq-EMwithC}) that only quantities formed from inner products of the
$(C^{\nu})$ vectors can have
physical consequences. The outer indices parameterize ten of these inner
products; these are the ten physical parameters and precisely the ten constants
that define the bilinear form in (\ref{eq-EMwithC}). So in essence, only the
magnitudes and relative lengths of these vectors are physical. The overall
orientation of the cluster of vectors has no physical consequences, and $SO(3,1)$
rotations of the entire cluster parameterize the six unphysical parameters.

To leading order, the four $f^{\mu}$ coefficients are just the angles that
parameterize a rotation in spinor space. It is natural then that $f$ is not
renormalized at this order; any radiative corrections to $f$ would actually be
quantum corrections to the Dirac matrix representation. Conversely, the choice
of Dirac matrices should not affect the renormalization of any of the theory's
other parameters. So there are no ${\cal O}
(f)$ terms in any of the $\beta$-functions of Lorentz-violating
QED~\cite{ref-kost4}, for example. At second order in $f$, on the other hand,
there are radiative corrections to $c$.

We have discussed a field redefinition that eliminates $f$ from the Lagrangian.
However, another type of field redefinition is
often used when the theory is considered
in the Hamiltonian framework (e.g. in~\cite{ref-bluhm1,ref-bluhm2,ref-bluhm3}). 
If $\Gamma^{0}$ is invertible, then the Dirac
equation may be recast in the Schr\"{o}dinger-like form
\begin{equation}
\label{eq-Hbad}
i\partial_{0}\psi=\left(\Gamma^{0}\right)^{-1}\left(i\vec{\Gamma}\cdot\vec{\nabla}
+m\right)\psi.
\end{equation}
However, the operator appearing on the right-hand side of (\ref{eq-Hbad}) will
not generally be Hermitian, because there were nonstandard time derivative terms
in the original Lagrangian. Using a field redefinition $\psi=
\left(\gamma^{0}\Gamma^{0}\right)^{-1/2}\psi'$, we may transform (\ref{eq-Hbad})
into a new equation with a Hermitian Hamiltonian, provided that $\gamma^{0}
\Gamma^{0}$ is positive definite~\cite{ref-kost3,ref-lehnert3}.

We shall now examine how these alternate field redefinitions
behave in the presence of a Lorentz-violating $f$ only, so that
$\Gamma^{0}=\gamma^{0}+if^{0}\gamma_{5}$. Invertibility of this matrix requires
only that $f_{0}^{2}\neq 1$. However, $\gamma^{0}\Gamma^{0}$ will not be
positive definite unless the stronger condition $f_{0}^{2}<1$ is met. This
condition for the existence of the field redefinition is also stronger than
the condition $f^{2}<1$ that we encountered when looking at transformations of
the Lagrangian---a fact which should be unsurprising.
In order to have a well-defined
Hamiltonian formulation, we must also be able to define the theory properly via
its Lagrangian. However, since the Hamiltonian method chooses a particular frame,
it can be less advantageous. The cost of choosing a reference frame in which
$\left|f^{0}\right|$ is greater than its minimum value $\sqrt{\max(f^{2},0)}$ is
that we must have $f_{0}^{2}<1$, rather than merely $f^{2}<1$, in order to define
the theory. In essence, by examining the theory in an inopportune frame,
we are not making the best use of the spacelike Lorentz-violating coefficients,
which could be used to improve the theory's behavior. Finally, we point out that
$f_{0}^{2}<1$, as it is not a Lorentz-invariant condition, could be violated,
even for small $f^{2}$, in a highly boosted frame; and this again illustrates
that the problems in defining the Hamiltonian are associated with choosing a poor
choice of frame when quantizing the theory.

While $f$ is unnecessary for our description of the SME fermion sector,
it is still possible that it might prove convenient to use this parameter in
specific situations. Effects that depend on $c$ in a
particular fashion might be more simply expressed in terms of $f$. For example,
some of
the most stringent bounds on $c$ for the electron come from observations of
synchrotron radiation from the Crab nebula~\cite{ref-altschul3}. The spacelike
coefficients so bounded
take the form $c_{jk}\hat{e}_{j}\hat{e}_{k}$, where $\hat{e}$ is a unit vector. So
this constraint is on exactly that part of $c$ that has the form $c_{jk}=\pm v_{j}
v_{k}$ for some vector $\vec{v}$. This suggests that a formulation in terms of
$f$ might be more succinct. Yet unfortunately, the bound in~\cite{ref-altschul3}
is one-sided. A positive $c_{jk}\hat{e}_{j}\hat{e}_{k}$ leads to a maximum
electron velocity
in the direction of $\hat{e}$, and that is a phenomenon with
readily measurable effects.
However, a negative $c_{jk}\hat{e}_{j}\hat{e}_{k}$
does not lead to a maximum velocity, so
no $c_{jk}=-\frac{1}{2}f_{j}f_{k}$ is excluded by this measurement---although, if
a bound on a negative $c_{jk}\hat{e}_{j}\hat{e}_{k}$
were available, it would immediately
translate into a bound on $|f_{j}\hat{e}_{j}|$ in a formulation of the theory
involving $f$.

Nothing that we have discussed will change if the conserved vector current is
coupled to a gauge
field. The derivative $\partial_{\mu}$ is simply replaced by a covariant
derivative $D_{\mu}$. The inclusion of the vector potential does not affect the
field redefinition in any way. This may seem a trivial observation, but there are
situations where similar conclusions do not hold. A $b$ term may be eliminated
from a massless noninteracting theory
by a different kind of field redefinition, $\psi'=e^{-i\gamma_{5}b^{\mu}x_{\mu}}
\psi$. This corresponds to separate translations
of momentum space for the left- and right-handed fermions. While an Abelian vector
coupling
does not appear to mix the two helicities, it is well known that chiral
symmetry is broken at ${\cal O}(\hbar)$ by the anomaly. So, even though it looks
like this field redefinition should eliminate $b$ entirely from the physics,
that coefficient can still contribute to quantum corrections.

Nor do we expect a coupling to gravity to affect our ability to eliminate $f$.
The field redefinition that transforms away this coefficient is really just
a change in the basis used for the Dirac matrices. A gravitational interaction is
not coupled in any way to the specific Dirac matrices used to define a theory, so
a rotation in spinor space is still allowed, even in curved spacetime.  This is in
contrast to what happens with the $a$ term, which cannot generally be removed
when there is a nontrivial spacetime background. The reason is that the field
redefinition $\psi'=e^{ia^{\mu}x_{\mu}}\psi$ which removes $a$ is $x$-dependent,
and this dependence interacts nontrivially with the covariant derivative.
However, since $f$ is removed by an $x$-independent field redefinition, there
are no analogous problems associated with its elimination.

Finally, we must address the issue of couplings to chiral gauge theories. As
previously noted, an $f$ term---but not a $c$ term---will mix left- and
right-chiral fermion fields. This appears to contradict the fact that a $f$ may
be converted into a $c$. However, the reasoning that leads to the contradiction
is actually based on an erroneous assumption. We have assumed that the chiral
projectors that appear in the Lagrangian should have the form $\frac{1\pm
\gamma_{5}}{2}$. However, while the chiral current $\bar{\psi}\Gamma^{\mu}
\gamma_{5}\psi$ is not conserved if $M=0$ and $f\neq0$, there is another
conserved current, $\bar{\psi}\Gamma^{\mu}\gamma'_{5}\psi$, with $\gamma'_{5}=
\gamma_{5}+{\cal O}(f)$. In fact, the necessary $\gamma_{5}'$ is simply
$-i\gamma'_{0}\gamma'_{1}\gamma'_{2}\gamma'_{3}=
e^{\frac{i}{2}f^{\mu}\gamma_{\mu}\gamma_{5}G\left(-f^{2}\right)}\gamma_{5}
e^{-\frac{i}{2}f^{\mu}\gamma_{\mu}\gamma_{5}G\left(-f^{2}\right)}$.
The modified chiral current is conserved, because
$\gamma'_{5}$ anticommutes with all
the $\Gamma^{\mu}$. So the theory can consistently be coupled to an
$SU(2)_{L}$ gauge group, provided the left-chiral projector used is actually
$\frac{1-\gamma'_{5}}{2}$. (In a similar vein, there is a modified $CPT$ operator
under which the theory with $f$ is even, like the theory with $c$.)

However, the existence of a modified chirality operator is a special property of
the theory containing $f$. Such an operator does not exist for general
$\Gamma^{\mu}$. The question of whether such an operator exists is closely
tied to the relationship between $e$ and $f$, which we shall now briefly
discuss. For $m=m_{5}=0$, the energy-momentum relations have
the same form
for theories with either $e$ or $f$ as the sole Lorentz-violating coefficients.
(More generally, $e$ gives the same dispersion relation as theory with particular
$a$, $c$, and modified $m$ coefficients.)
Since the energy-momentum relations and particle statistics completely define
a noninteracting quantum field theory, this means the theories with either
solely $e$ or solely $f$ (and no masses) describe the same physics.
However, in a theory with an $e^{0}$ as its only form
of Lorentz violation, there is no matrix $\gamma_{5}+{\cal O}(e)$ that
anticommutes with $\Gamma^{0}=\gamma^{0}+e^{0}$. Yet
actually there is a field redefinition that will convert an $e$ term into an
$f$ term in precisely the $m=m_{5}=f=0$ case:
\begin{equation}
\label{eq-eeliminate}
\psi'=e^{-i\frac{\pi}{4}\gamma_{5}}
\psi=\frac{1}{\sqrt{2}}(1-i\gamma_{5})\psi.
\end{equation}
The terms with $e$ and $f$ have the
same Dirac matrix structures as $m$ and $m_{5}$, respectively, so it might seem
obvious that in the massless theory,
we can eliminate $e$ in favor of $f$, just as $m$ could be eliminated in
favor of $m_{5}$. However, because the full term containing $e$ involves a
derivative, the necessary field redefinition is nonlocal. The only exception to
this is if $f=0$ initially, so that the argument of
the inverse tangent in the analogue of (\ref{eq-m5}) is singular; the resulting
transformation is exactly (\ref{eq-eeliminate}). The necessary field redefinition
does not vanish as $e\rightarrow0$, and so the theory's modified
chirality operator does not have the form $\gamma_{5}+{\cal O}(e)$. If the
theory initially contains both nonzero $e$ and $f$, then we could
still attempt to construct a new chirality operator
via a field redefinition. The resulting operator would formally obey the correct
Clifford algebra anticommutation relations, but it would actually be nonlocal.
In the presence of interactions with additional spacetime-dependent fields,
the nonlocal field redefinitions will not work to eliminate $e$ from the
theory, because $\partial^{\mu}$ and $x_{\mu}$ do not commute.
Therefore, unless either $m=m_{5}=f=0$ or $m=m_{5}=e=0$, the
explicit breaking of chiral symmetry is real and unavoidable. Moreover, if
$m\neq0$, then $e$ and $f$ are definitely not equivalent. There can be
physical effects of ${\cal O}(e)$ involving gravity, while $f$ can always be
eliminated in favor of a $c$ that is ${\cal O}(f^{2})$.

So we have seen that the $f$ coupling is really quite special. While it
has no effects
at linear order, it is not trivial in general. However, there is no unique
phenomenology associated with this form of Lorentz violation. The $f$ coefficient
can be removed from the theory by a spacetime-independent
field redefinition, which replaces a pure $f$ term with a $c$ term,
provided that $f^{2}<1$. However, only small values of $f^{2}$ are really
interesting, both because that
represents the only possible physical regime, and because
there are causality violations at an unacceptably low scale if $f^{2}$ is
comparable to unity. For small $f$, the $f^{\mu}$ coefficient
is equivalent to a $c^{\nu\mu}\approx-\frac{1}{2}f^{\nu}f^{\mu}$.

The field redefinition that eliminates $f$ in favor of $c$ is compatible with
vector and chiral gauge couplings, as well as a coupling to gravity. This
implies that $f$ is actually a
completely extraneous parameter in the SME. For each fermion
species, it may be transformed away. So further consideration of
the $f$ coefficients is unnecessary, and this
represents an important simplification to the structure of
Lorentz-violating effective field theory.

\section*{Acknowledgments}
The author is grateful to V. A. Kosteleck\'{y} for helpful discussions.
This work is supported in part by funds provided by the U. S.
Department of Energy (D.O.E.) under cooperative research agreement
DE-FG02-91ER40661.

\end{document}